\documentstyle{disk}
\input{epsf.sty}

\def\plotone#1{\centering \leavevmode \epsfxsize=0.99\columnwidth \epsfbox{#1}}

\begin{document}

\title{%
Three-dimensional SPH simulations of \\ accretion discs
\footnote{To be published in {\it Disk Instabilities in  Close Binary Systems
-- 25 Years of the Disk-Instability Model}, Mineshige S. \& Wheeler J.C., eds, Universal
Academy Press, 1999}
}


\author{Henri M.J.\ BOFFIN\\
{\it Royal Observatory of Belgium,\\ 
3 av. Circulaire, B-1180 Brussels (Henri.Boffin@oma.be)}\\
Kei HARAGUCHI and Takuya MATSUDA\\
{\it Department of Earth and Planetary Sciences,\\
Kobe University, Rokkoudai machi, 657-8501 Kobe, Japan}}

\maketitle

\section*{Abstract}
We discuss some 3D numerical simulations of accretion discs using the
SPH method and a polytropic equation of state.
We show that discs exist even for as large value of the polytropic index as
1.2, and that these discs are always in hydrostatic balance.
We also show that even without any inflow, spiral shocks appear in the
discs.

\section{Introduction}

Numerical studies of accretion discs have been mostly restricted
to 2D cases, due to computing time limitations. Among many things, these 2D
simulations have shown that spiral shocks appear in inviscid discs
(e.g. Sawada et al. 1987).
Recently some 3D simulations have been carried out (see Yukawa, Boffin \& 
Matsuda, 1997 for an uncomplete list), mostly using particles methods.
These simulations were apparently unable to generate spiral shocks
in the accretion disc, but this could be related to the fact that 
they used either an isothermal or pseudo-isothermal equation of state, either
neglected pressure effects or used too low resolution. 
We have run three-dimensional Smoothed Particle Hydrodynamics (SPH; see e.g.
Monaghan 1992 for a review) simulations
with a polytropic equation of state. This method includes self-consistently the
effect of pressure forces and we checked that we could always resolve the
disc in the vertical dimension. Concerning this last point, we therefore used
a variable smoothing length (which, in SPH, decides the resolution) and checked that
at each point in space, the smoothing length, $h$, was smaller than the disc scale
height, $H$. 
Details of the method and of some of the results can be found in Yukawa et al. (1997).

\begin{figure}[htbp]
    \plotone{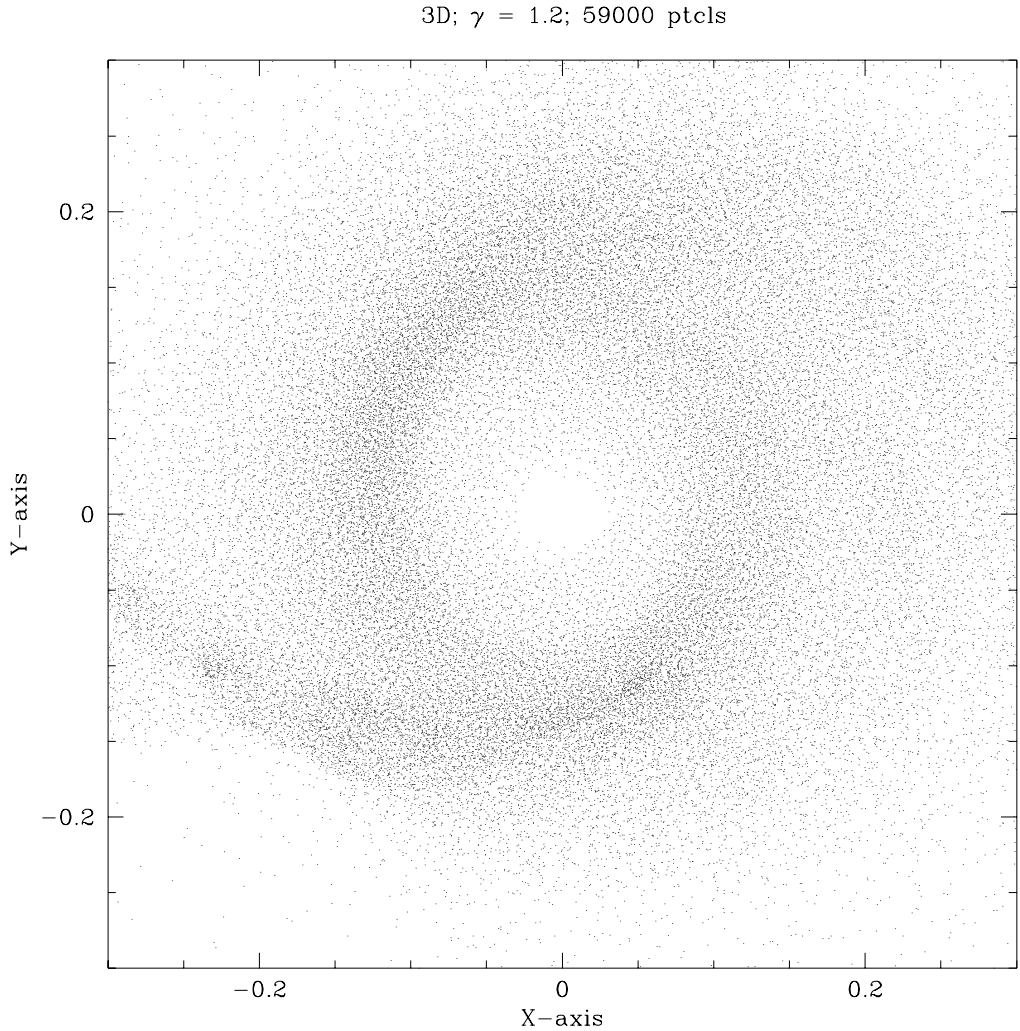}
   \caption{\label{BHM:slice2} Structure of the flow in the orbital plane, for our 
simulation with a mass ratio of 1.0 and $\gamma=1.2$, after 2 orbital periods. Only the region close to the
primary is shown. A total of 59,000 particles was used.} 
\end{figure}

\section{Results : mass inflow}

In figure \ref{BHM:slice2}, we show the flow at the end (i.e. two orbital periods)
of our simulation with mass inflow when we use a polytropic index, $\gamma$=1.2 .
As can be seen, a spiral structure is clearly present, confirming the fact that
SPH is able to tracks these structures but, more importantly, that these structures
are present in 3D accretion flows.
This result also confirms that a disc does form in 3D, even for such a large value of the
polytropic index. Moreover, the disc is in hydrostatic balance, as its disc height
is precisely equal to the value expected:
$H \simeq {c_s}/{\Omega}$,
where $c_s$ is the sound speed and $\Omega$ is the angular velocity.
Because, we use a rather large sound speed as initial condition (0.1, where the orbital
velocity corresponds to 1.0) and a large polytropic index, the disc we obtain is rather
hot, hence rather thick ($\frac{H}{r} \simeq 0.2-0.3$). 
For the smaller value of $\gamma$ used, 1.1 and 1.01, we obtain smaller disc heights : 
0.12 to 0.2 and 0.09, respectively. In both cases, the hydrostatic balance in the vertical 
direction holds true.
And in all cases, the ratio between the local vertical disc height (i.e. the disc semi-thickness) and the
local smoothing length lies between about 2 and 6. Thus, we have certainly resolved the disc
vertically.

\begin{figure}
   \plotone{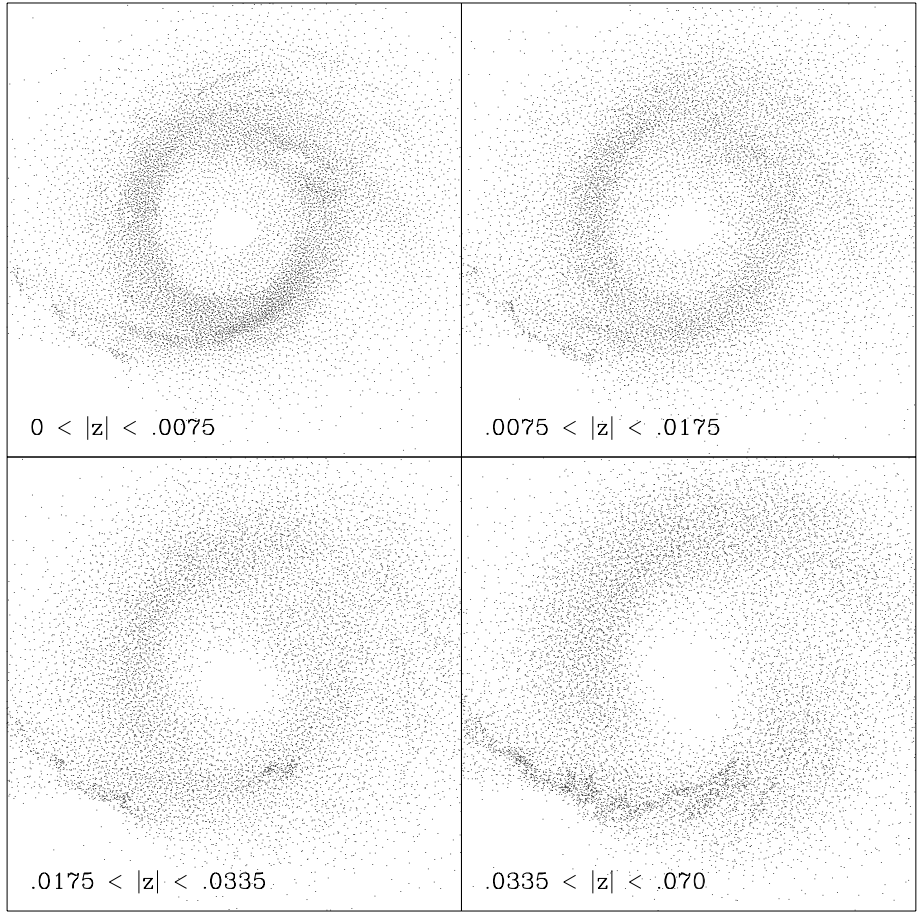}
   \caption{\label{BHM:slice} Same as fig.\ref{BHM:slice2}, except that we have subdivided
the particles following their height, $z$, above the orbital plane, in such a way that there are as
many particles (13,200) in each frame : 
\hfill \break \mbox{\hspace{6.0cm}} a) $0.0000 < z < 0.0075 $,
\hfill \break \mbox{\hspace{6.0cm}} b) $0.0075 < z < 0.0175 $, 
\hfill \break \mbox{\hspace{6.0cm}} c) $0.0175 < z < 0.0335 $, 
\hfill \break \mbox{\hspace{6.0cm}} d) $0.0335 < z < 0.0700 $.} 
\end{figure}

Just a note in passing concerning the viscosity present in our code.
We use the standard artificial viscosity of SPH which, as shown e.g. by Murray (1996), has
an equivalent shear viscosity, $ \nu \simeq 0.1 ~ \alpha_{\rm SPH}  c_s  h$. 
In term of the Shakura-Sunyaev $\alpha$-viscosity, $\alpha_{\rm SS}$, this can be rewritten, 
\begin{equation}
\alpha_{\rm SS} \simeq \frac{\alpha_{\rm SPH}}{10} \frac{\Omega ~ h}{c_s} 
\simeq \frac{\alpha_{\rm SPH}}{10} \frac{h}{H}.
\end{equation}
With the value of $\alpha_{\rm SPH}=1$ used, we therefore have an equivalent 
$\alpha_{\rm SS}$ of 0.02 to 0.05.

\begin{figure}
   \plotone{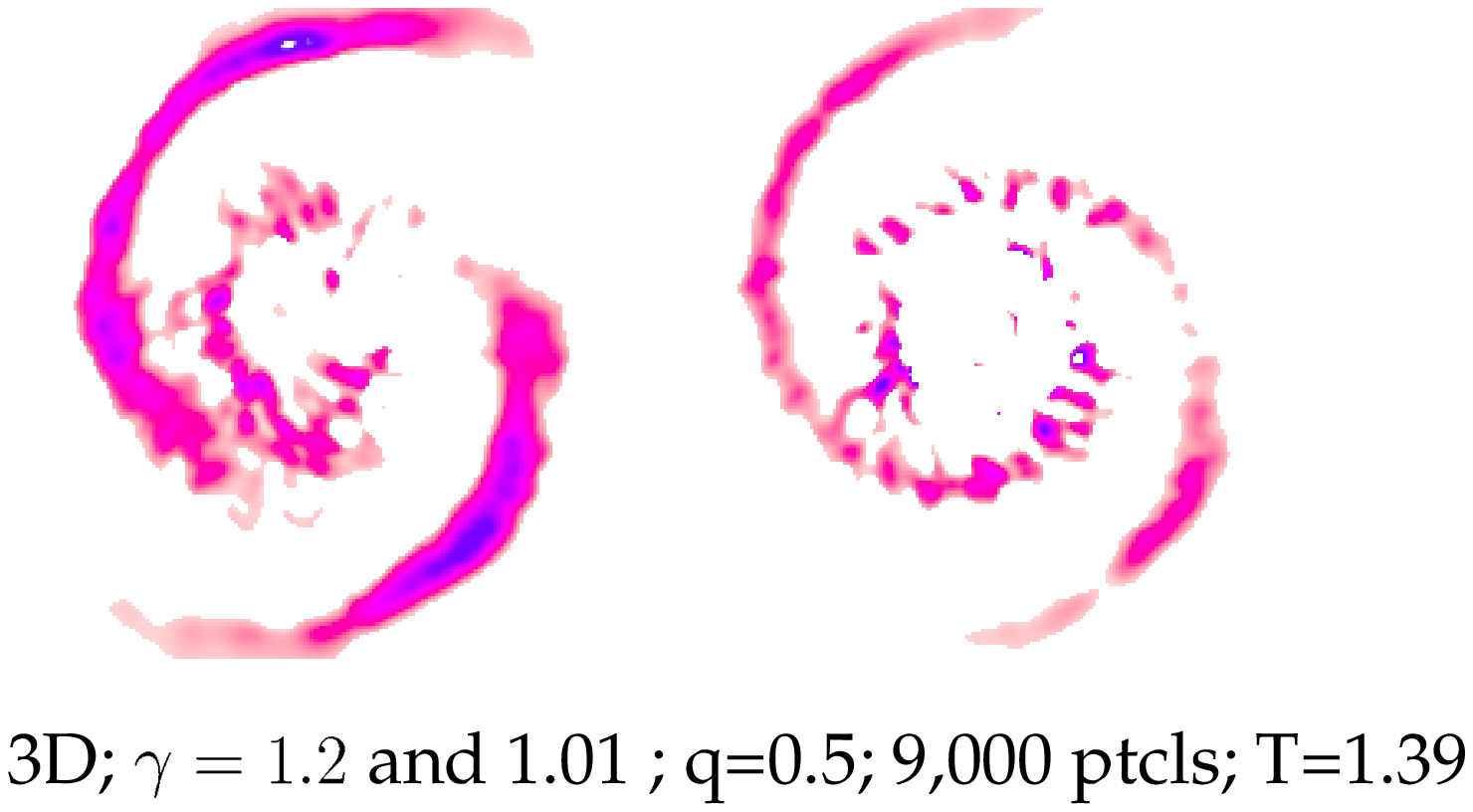}\\
   \plotone{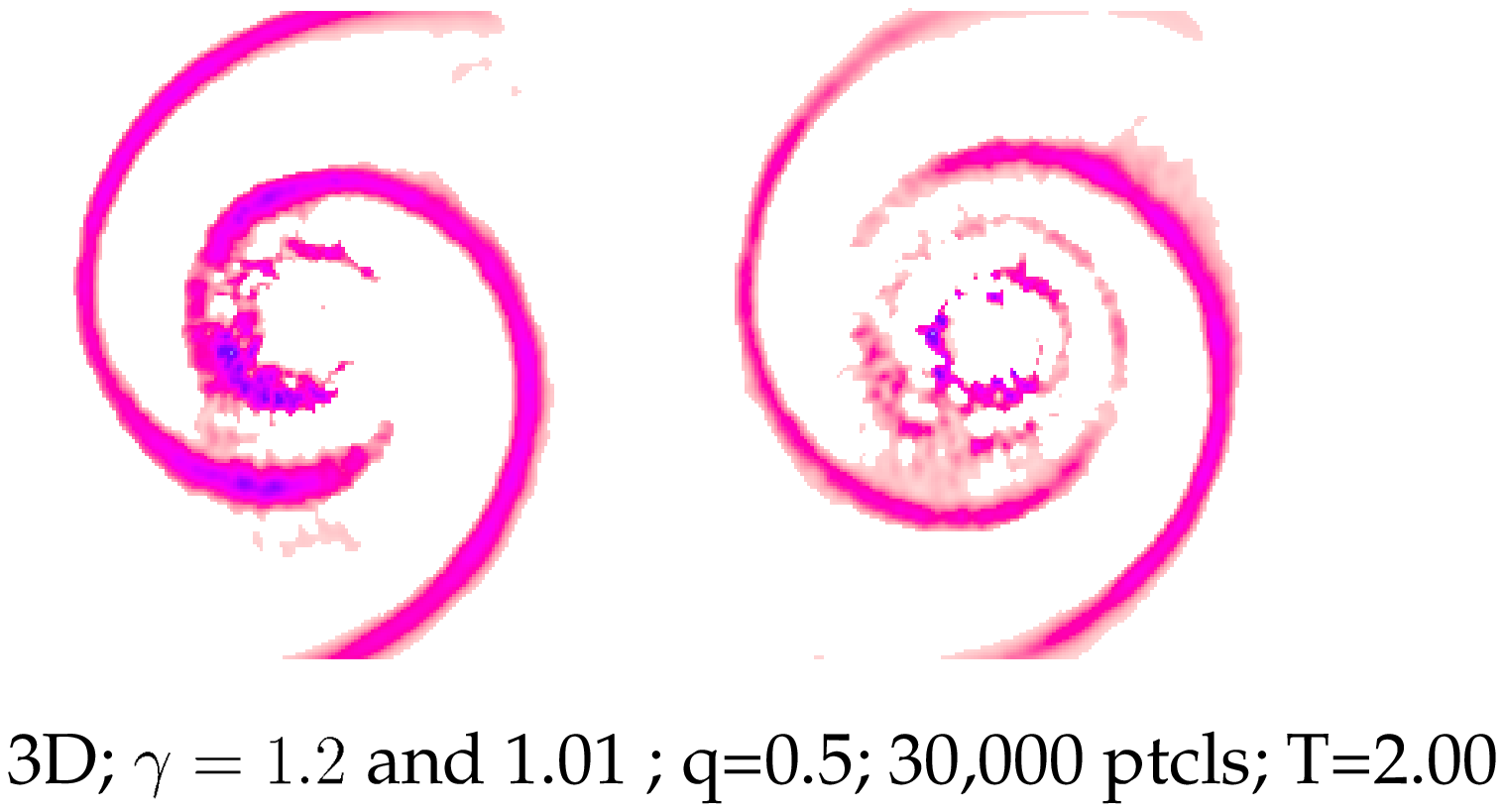}
   \caption{\label{BHM:comp} Results of our simulations without mass inflow for two value of
the polytropic index, $\gamma$=1.2 (left) and 1.01 (right). The upper frames
correspond to a lower resolution (9000 particles) than the lower frames (30 000
particles). The mass ratio is 0.5 and the plots represent greyscale maps of the asymmetric part 
of the density field. Note the effect of increasing the resolution.} 
\end{figure}

It has to be noted that we cannot claim to have obtained a true steady 
state as the mass in the disc is still increasing at the end of the simulations. Two-dimensional
simulations (Boffin et al., in preparation) show us that several tens of orbital periods
are necesary to reach a steady state. However, in our 3D simulations, we can see that the
structure of the flow does not change after, say, one orbital period. We therefore believe that we have 
reached a "quasi-steady state" and can study the final structure of the flow.
We cannot, however, make any definite claims about the mass accretion rate.
From figure \ref{BHM:slice2}, we also observe that we do not have a true "hot spot" but
more a kind of "hot line". This is, we believe, again due to the large initial sound speed,
resulting in a very wide inner Lagrangian stream.

In figure \ref{BHM:slice}, we show the same as in figure \ref{BHM:slice2}, except
that we have divided the particles following their height above the orbital plane.
This can be used to study the possible variation of the disc height with the orbital phase
as obtained by Hirose et al. (1991). We do not seem to find any conclusive variations, however.
Also, we cannot observe any stream overflow in the z-direction as obtained by Armitage \& Livio
(1996). The reason for this discrepancy is unclear and we are presently working on this.
Possible reasons are : their use of a large viscosity, their initial conditions, 
our large initial sound speed, ...

\section{Results without mass inflow}

We have also performed several simulations without any mass inflow. In this case,
a disc is initially set-up around the primary, so that it is resolved vertically
and in hydrostatic balance. It is then evolved with the full potential of the
binary system taken into account.
Here again, as shown in figure \ref{BHM:comp}, which is a greyscale map of the
asymmetric component of the density, spiral shocks can clearly be seen, 
both in the $\gamma$=1.2 and $\gamma$=1.01 cases. Thus, these spiral shocks are not the result
of the inner Lagrangian flow. This is not a surprise if, as believed, the spiral 
structures are due to the tidal force of the companion ({\it e.g.} Savonije et al. 1994).

Figure \ref{BHM:comp} also shows the importance of resolution : although with 9,000 particles
we cannot find any severe difference between $\gamma$=1.2 and 1.01, 
this is no more true with 30,000 particles. For $\gamma$=1.01 indeed, in the inner part 
of the disc, the spirals become more tightly wound, a result well known in 2D 
({\it e.g.} Sawada et al. 1987). The reason for this difference may lie in the fact that 
for the $\gamma$=1.2 case, the Mach number of the flow always remains smaller than 10,
while for the $\gamma$=1.01 case, it starts at a little below 10 in the outer part
of the disc to reach above 30 in the inner part. 
It was already shown by, {\it e.g.}, Savonije et al. (1994) that the higher the Mach number, 
the more tightly wound the spiral. What is noticeable in our 3D simulations, 
is the fact that we cannot make any clear distinction between a cooler disc 
($\gamma$=1.01) and a hotter disc ($\gamma$=1.2) when we restrict ourselves to the
outer parts of the disc. If confirmed, this result may be useful to reproduce the
observations of spiral shocks in IP Peg (Steeghs et al. 1997) even when we deal with 
the cool disc typical of a dwarf nova.

\section{References}

\vspace{1pc}

\re
1.\ Armitage P.J., Livio M.\ 1996, ApJ 470, 1024
\re
2.\ Hirose M., Osaki Y., Mineshige S.\ 1991, PASJ 43, 809
\re
3.\ Monaghan J.J.\ 1992, Ann.Rev.Astr.Astrophys. 30, 543
\re
4 .\ Murray J.R.\ 1996, MNRAS 279, 402
\re
5.\ Savonije G.J., Papaloizou J.C.B., Lin D.N.C.\ 1994, MNRAS 268, 13
\re
6.\ Sawada K., Matsuda T., Inoue M., Hachisu I.\ 1987, MNRAS 224, 307
\re
7.\ Steeghs D., Harlaftis E.T., Horne K.\ 1997, MNRAS 290, L28
\re
8.\ Yukawa H., Boffin H.M.J., Matsuda T.\ 1997, MNRAS 282, 321

\end{document}